\documentclass[runningheads]{llncs}
\usepackage{amsmath}
\usepackage{amssymb}
\usepackage{nccmath}
\usepackage{mathtools}
\usepackage{xspace}
\usepackage{enumitem}
\usepackage{xcolor}
\usepackage{algorithm}
\usepackage[noend]{algorithmic}

\DeclareMathOperator*{\argmax}{arg\,max}

\newtheorem{observation}{Observation}
\newcommand{\pref}{\succ}

\newcommand{\F}{\ensuremath{\mathcal{F}} \xspace}
\newcommand{\T}{\ensuremath{\mathcal{T}} \xspace}

\begin{document}

\title{Democratic Forking: \\ Choosing Sides with Social Choice}

\author{Ben Abramowitz\inst{1} \and Edith Elkind\inst{2} \and Davide Grossi\inst{3} \and Ehud Shapiro\inst{4} \and Nimrod Talmon\inst{5}}

\institute{Rensselaer Polytechnic Institute, Troy NY 12180, USA \email{abramb@rpi.edu} \and University of Oxford, Oxford, United Kingdom \email{eelkind@gmail.com} \and University of Groningen and University of Amsterdam, Groningen and Amsterdam, The Netherlands \email{d.grossi@rug.nl} \and Weizmann Institute of Science, Rehovot, Israel \email{ehud.shapiro@weizmann.ac.il} \and Ben-Gurion University, Be'er Sheva, Israel \email{talmonn@bgu.ac.il}}

\authorrunning{B. Abramowitz et al.}

\maketitle

\begin{abstract}
Any community in which membership is voluntary may eventually break apart, or fork.
For example, forks may occur in political parties, business partnerships, social groups, and cryptocurrencies.
%
Forking may be the product of informal social processes or the organized action of an aggrieved minority or an oppressive majority.
%
The aim of this paper is to provide a social choice framework in which agents can report preferences not only over a set of alternatives, but also over the possible forks that may occur in the face of disagreement.
We study the resulting social choice setting, concentrating on stability issues, preference elicitation and strategy-proofness.
%
\end{abstract}


\section{Introduction}
Collective decisions can produce conflict when no outcome is acceptable to all agents involved. Tensions may arise while options are being considered and discussed, but they may also appear or worsen once a decision has been made, particularly if preference strengths have not been accounted for or if some agents are disenfranchised entirely. 

In many situations there is an implicit recourse for the frustrated and the downtrodden---they can leave. An agent, or a set of agents, may leave a community if they are sufficiently dissatisfied with the outcome of a decision, or a bundle of decisions. 
When group cohesiveness is valued, it is sensible for decision makers to consider how each potential decision might effect camaraderie. This is commonly done informally through discussion when group sizes are small enough to deliberate, and through polling when communities are large. Here we propose a formal mechanism to take such considerations into account.

In typical voting scenarios, agents express preferences over alternatives and a single alternative is elected, which must then be universally accepted as the outcome by all agents in the community. In our setting, agents do not have to accept a particular alternative, and can fork instead. Consequently, the set of possible outcomes for every decision is not just the set of alternatives; rather, an outcome is a partition of the agents into one or more coalitions, with each coalition selecting an alternative that appeals to its members.

By designing voting rules that account for forking preferences, we empower aggrieved minorities to leverage the threat of leaving in order to pressure the majority into concessions. Importantly, we enable minorities to coordinate 
with minimal overhead, by eliciting additional information during the voting process.
In our model, voters individually indicate conditions under which they prefer to leave,
and the mechanism then identifies a group that can benefit from forking, thereby 
eliminating the need for campaigning or coordination among the disgruntled minority.


The value of stability is inherent in digital and analogue communities alike.
In most current blockchain protocols (i.e., proof-of-work and proof-of-stake) a fork can only be initiated by a majority or a powerful minority.
When a fork does occur, enacted by only a subset of the agents, all agents must determine what ``side" of the fork they want to be on.\footnote{In blockchain forks, currency owners may be able to run both protocols, but miners must choose how to allocate their computing resources, and programmers must choose how to allocate their personal time.}
%
%
In this sense, the forks we study are also relevant to version control systems such as Git, where anyone can initiate a fork. In our setting this is an edge case where a voter is willing to split from the group by themselves and others may follow, but we also allow groups of voters to fork together when none of them may be willing to do so on their own. 

Since forks can be tumultuous, we wish to design democratic systems that enable communities to efficiently find states that are stable, in the sense that no further forks will occur. To this end, we put forward a formal model that approaches this challenge from the perspective of computational social choice.

\subsubsection*{Related Work}

Our paper is positioned at the interface of the research on blockchain technology and computational social choice.

\subsubsection*{Blockchain}
A blockchain is a replicated data structure designed to guarantee the integrity of data (e.g., monetary transactions in cryptocurrency applications such as Bitcoin~\cite{bitcoin}) and computations on them, combined with consensus protocols, which allow peers to agree on their content (e.g., who has been paid) and to ensure that no double spending of currency has occurred (see \cite{narayanan16handbook,wattenhofer17distributed} for recent overviews). By now, blockchain is an established technology, and cryptocurrency applications are attracting considerable attention~\cite{cryptocurrency,cryptoimpact}.


When the community of a specific blockchain protocol---such as the Bitcoin protocol---is not satisfied with it, it may break into several subcommunities. The community typically consists of developers, who build the software, miners, who operate the protocol, and users, e.g., account holders.  When some of the developers of the current protocol decide to modify it, they create an alternative branch that obeys their new protocol, and if it attracts a sufficient number of miners and users, the result is a so-called \emph{hard fork}. Several such hard forks have been documented, including among the most influential cryptocurrencies. Bitcoin (cf.~\cite{hardfork}), despite its relatively short history, has already undergone seven hard forks.
At the moment, a key feature of these hard forks is that they happen through an informal social process, and, crucially, in ways that are completely exogenous to the protocol underpinning the blockchain. In blockchain terminology, they are said to happen `off-chain'. This points to a lack of governance in most current blockchain systems, for better or worse.

Against this backdrop there have been attempts at incorporating protocol amendment procedures within blockchain protocols themselves (so-called `on-chain' governance, cf. \cite{tezos}); more generally, the issue of governance is attracting increasing attention~\cite{beck2018governance,reijers2018now}.
We are not aware, however, of research that approaches forking as a social choice problem, and aims for an algorithmic solution. We lay the foundations for this approach here.

\subsubsection*{Computational social choice}
Social choice theory studies preference aggregation methods for various settings~\cite{brandt2016handbook}. Our social choice setting is closely related to assignment problems, as the result of a fork is that each agent is ``assigned'' to a community.
%
In this context, we mention works on judgement aggregation~\cite{Grossi_2014},~\cite[ch. 17]{brandt2016handbook} (which, formally, can capture assignments as well) and on partition aggregation~\cite{bulteaupartition}. 
To the best of our knowledge, the specific social choice setting we consider is novel.
In a broader context, we mention work on stability in coalition formation games~\cite{cgt-book},~\cite[ch. 15]{brandt2016handbook} as well as the recent paper on deliberative majorities~\cite{deliberativemajorities}, which studies coalition formation in a general voting setting.
Our model is also related to the group activity selection problem with (increasing) ordinal preferences ({\sf o-GASP}) \cite{darmann2015group} (see also \cite{darmann2018group,darmann2018social}), where our notion of stability corresponds to core
stability in {\sf o-GASP}. However, due to our focus on strategy-proofness, and the fact that our setting does not admit a no-choice option (void activity in {\sf o-GASP}), most of the existing
results for {\sf o-GASP} are not directly relevant to our study, so we chose not to use 
the {\sf o-GASP} formalism.

\subsubsection*{Version control}
Forking is not limited to the cryptocurrency setting; in particular, forking is relevant to projects of open-source code, in which a community jointly writes a piece of code and may experience different opinions regarding the code that is being written.
Indeed, there is some work on using social choice mechanisms (and, in particular liquid democracy) for revision control systems~\cite{liquidgit}. Others have been studying the phenomena of forking in open source projects; see, e.g., the work of Zhou et al.~\cite{zhou2020has}.

\subsubsection*{Outline and Contributions}

We describe a formal model of social choice for community forking, in which agents report their preferences over alternatives relative to possible forks. Throughout the paper, we focus on the setting where the number of available alternatives is two; towards the end of the paper, we discuss the challenges in extending our approach to three or more alternatives.
The paper is structured as follows.
Section~\ref{section:model} describes the formal model.
Section~\ref{section:stability} examines whether stable solutions always exist and whether they can be found efficiently in terms of computation and elicitation.
In Section~\ref{section:strategicbehavior}, we consider strategic agent behavior.
In Section~\ref{sec:three} we discuss the extension of our framework to more than two alternatives. We conclude in Section~\ref{sec:conclusions}.
The main contributions of our paper are as follows:
\begin{itemize}

\item 
We devise a polynomial time algorithm (Algorithm~\ref{algorithm:general}) for our setting that finds a stable assignment for a very broad and natural domain restriction.

\item
We propose a modification of Algorithm~\ref{algorithm:general} (Algorithm~\ref{algorithm:generaliterative}) that allows for efficient iterative preference elicitation.

\item
We prove an impossibility result (Theorem~\ref{theorem:notstrategyproof}), showing that there is no algorithm that is strategyproof for profiles with more than one stable assignment.

\item
We establish that Algorithm~\ref{algorithm:general} is strategyproof for profiles that admit a unique stable solution; the impossibility result mentioned above then implies that it is optimal in that sense.

\end{itemize}


\section{Formal Model}\label{section:model}

\subsubsection*{Setting}
We have a set of agents $V = \{v_1, \ldots, v_n\}$. This community will vote on a set of two alternatives $\{A,B\}$ (say, cryptocurrency protocols or locations).
However, unlike in most voting scenarios, the agents are not all bound to accept the same winning alternative.
Agents have the ability to \emph{fork}, or forge a new community centered around the ``losing" alternative.
Ultimately, either all of the agents will remain in a single community or they will split into two communities that have accepted opposite alternatives.
%

\subsubsection*{Agent Preferences}

Agents care about what alternative their community adopts and how many people are in their community, but not the identities of the other agents in their community.
We can represent agent preferences as total orders over the possible tuples $(S,j)$, where $S \in \{A,B\}$ is the alternative to which they are assigned and $j \in [1,n]$ is the number of agents in their community (including themselves and $j-1$ other agents). We denote the set of all such tuples by $\mathcal{S}$
The preference relation $(A,j) \pref_i (B,k)$ means that agent $v_i$ would prefer to be in a community of size $j$ that accepts alternative $A$ rather than a community of size $k$ that accepts alternative $B$. 
Agent preferences are \emph{monotonic} in the size of their community, so given a fixed alternative, they would always prefer to be in a larger community.
Formally, for each agent $v_i \in V$ we have $(S,j) \pref_i (S,k)$ for all $1 \leq k < j \leq n$ and $S \in \{A,B\}$.
We denote the set of all monotonic total orders over $\mathcal{S}$ by $\T$.
For $S\in\{A, B\}$, let $V_S^*$ denote the set of agents who prefer $(S,n)$ to $(S', n)$. 
We will overload notation and use $v_i$ to represent both an agent $v_i \in V$ and their preference ordering $v_i \in \T$. In a similar fashion, $V$ is the set of agents and also the preference {\em profile}, or collection of the voters' total orders, $V \in \T^n$.
%
%
We refer to the pair $(V, \{A, B \})$, with $V \in \T^n$, as a {\em forking problem}.

\begin{example} \label{ex_three_agents}
Suppose we have $n = 3$ agents, $V = \{v_1, v_2, v_3\}$. Consider the preferences of a single agent $v_i$. By monotonicity, $(A, 3) \pref_i (A, 2) \pref_i (A, 1)$ and $(B, 3) \pref_i (B, 2) \pref_i (B, 1)$ must hold for all agents $v_i \in V$. However, these two orders may be interleaved differently for different agents.
\end{example}

\subsubsection*{Assignments}
We refer to the community that accepts alternative $A$ (resp., $B$) as community $A$ (resp., $B$).
An assignment $f: V \rightarrow \{A,B\}$ assigns agents to one of the two communities, and we denote by $f(v_i) \in \{A,B\}$ the community into which agent $v_i$ is placed.
The set $\F$ is the set of all $2^n$ possible assignments, or partitions, of the agents.
Voters' preferences over $\mathcal{S}$ induce preferences over assignments in $\F$: a voter
$v_i$ prefers an assignment $f$ to an assignment $g$ if
$(f(v_i), |f^{-1}(f(v_i)|) \pref_i (g(v_i), |g^{-1}(g(v_i)|)$.
Given a forking problem $(V, \{A, B \})$ a voting rule $R : \T^n \rightarrow \F$ selects an assignment $R(V) = f \in \F$.
We let $a = |f^{-1}(A)|$ be the size of community $A$, and similarly for $b = |f^{-1}(B)|$.
%


\section{Stability}\label{section:stability}
Our primary goal is to construct stable assignments.
An assignment is stable if no subset of agents has an incentive to move simultaneously to a new community.

\begin{definition}[$k$-Stability]
An assignment $f : V \to \{A, B\}$ is 
{\em stable} if there is no assignment $f' : V \to \{A, B\}$ 
such that each voter $v_i$ with $f'(v_i)\neq f(v_i)$ prefers $f'$ over $f$. 
\end{definition}
A voting rule $R$ is {\em stable} if it returns a stable assignment whenever one exists.

\begin{example}\label{example:stable}
Consider two agents, $V = \{v_1, v_2\}$, where $v_1: (A,2) \pref_1 (A,1) \pref_1 (B,2) \pref_1 (B,1)$ and $v_2: (B,2) \pref_2 (B,1) \pref_2 (A,2) \pref_2 (A,1)$. Each agent would prefer to be alone at their preferred alternative to being together with the other agent at their less preferred alternative.
Thus, the only stable assignment $f$ has $f(v_1) = A$ and $f(v_2) = B$.
\end{example}

\subsection{Finding Stable Solutions}

When preferences are monotonic, there must be at least one stable assignment, and it can be computed in polynomial time.

\begin{algorithm}[t]
\caption{General Stable Assignment Rule}
\label{algorithm:general}
\begin{algorithmic}
\STATE $V_A = V$, $V_B = \emptyset$, $a \gets |V_A|$, $b \gets |V_B|$

\WHILE{{\tt true}}
    \STATE $k \gets \max\{j: 0\le j\le a,  |\{v_i \in V : (B, b+j) \pref_i (A,a)\}| \geq j\}$
    \IF{$k = 0$} 
        \STATE return $\{V_A,V_B\}$
    \ELSE
        \STATE Let $X = \{v_i \in V : (B, b+k) \pref_i (A,a)\}$
        \STATE $V_B \gets V_B \cup X$, $V_A \gets V_A \setminus X$
        \STATE $a \gets |V_A|$, $b \gets |V_B|$
    \ENDIF
\ENDWHILE
\end{algorithmic}
\end{algorithm}

\begin{theorem} \label{theorem:generalstable}
  There is a polynomial time assignment rule (Algorithm \ref{algorithm:general}) that finds a stable assignment for any monotonic profile.
\end{theorem}

\begin{proof}
Consider Algorithm~\ref{algorithm:general}. Let $a = |V_A|$ and $b = |V_B|$. Initially, we place all agents in $V_A$, so $a = n$ and $b = 0$. 
If this assignment is not stable, then there exists a subset $X$ of some $k > 0$ agents that all prefer $(B,k)$ to $(A,n)$. We move all these agents to $V_B$.
Monotonicity implies that moving additional agents from $V_A$ to $V_B$ will never cause agents in $V_B$ to want to move back to $V_A$; thus, as long as we are not in a stable state, there must be a subset of agents at $V_A$ who would prefer to move together to $V_B$. As long as such a set of agents exists, we continue to move them over together.
This procedure halts in at most $n$ steps, and when it halts, the result must be stable, as there is no subset of agents who will move together. 
A naive implementation of the algorithm loops at most $n$ times, and each computation of the set $X$ of agents to move takes $O(n^2)$ time.
\qed\end{proof}

\subsection{Elicitation}

Algorithm~1 does not use all of the information in agents' preferences. An iterative version of the algorithm can ask only for the information it needs.
Instead of assuming that the total orders of all agents are given explicitly in the input, we place all of the agents at $A$ and at each iteration we ask the $a$ remaining agents at $A$ for the minimum value $j$ such that if $j$ agents could be moved to $B$, they would now prefer the new community $(B, b+j)$ over their current community $(A,a)$.
Once an agent has been moved to $B$ there is no need to ask them for any more information.
In Algorithm~\ref{algorithm:generaliterative} we repeatedly query agents about the conditions under which they are willing to leave their current community.
Agents indicate their preferences with a single integer that says how many agents would have to move with them for them to prefer leaving over the status quo. 

\begin{algorithm}[t]
\caption{General Stable Assignment Rule with Iterative Elicitation}
\label{algorithm:generaliterative}
\begin{algorithmic}
\STATE $V_A = V$, $V_B = \emptyset$, $a \gets |V_A|$, $b \gets |V_B|$

\WHILE{{\tt true}}
    \STATE Ask each agent $v_i$ in $V_A$ for the smallest value ${j \in [0,a]}$ such that $(B, b+j) \pref_i (A,a)$
    \STATE $k \gets \min\{j: j \in [0,a],  |\{v_i \in V : (B, b+j) \pref_i (A,a)\}| \geq j\}$
    \IF{$k = 0$}
        \STATE return $\{V_A,V_B\}$
    \ELSE
        \STATE Let $X = \{v_i \in V : (B, b+k) \pref_i (A,a)\}$
        \STATE $V_B \gets V_B \cup X$, $V_A \gets V_A \setminus X$
        \STATE $a \gets |V_A|$, $b \gets |V_B|$
    \ENDIF
\ENDWHILE
\end{algorithmic}
\end{algorithm}

Ideally, we would like to only have to query each agent a small number of times.
If agents' preferences are structured, it becomes possible to compute stable assignments with little information.
To capture this intuition, we introduce the concept of non-critically-interleaving preferences.

\begin{definition}[Non-critically-interleaving]
A preference is \emph{non-critically-interleaving} if it is monotonic and $(A,j) \pref (B,n) \pref (B, n-j) \pref (A, j-1)$ or $(B,j) \pref (A,n) \pref (A, n-j) \pref (B, j-1)$ for some $j \in [1,n]$.
A profile is \emph{non-critically-interleaving} if it contains only non-critically-interleaving preferences.
\end{definition}

When preferences are non-critically-interleaving, we only need to ask each agent whether they prefer $(A,n)$ or $(B,n)$, and the minimum value of $j$ such that they would rather be at their preferred alternative in a coalition of size $j$ than at the other alternative in a coalition of size $n$.
From this information the relevant part of the preference order of each agent can be inferred, and so Algorithm \ref{algorithm:generaliterative} will compute a stable assignment.

\subsection{Uniqueness}

While at least one stable assignment must exist for all monotonic profiles (Theorem~\ref{theorem:generalstable}), it is not necessarily unique. 

\begin{example}\label{example:manystable}
Let $V = \{v_1, v_2, v_3, v_4\}$ be a set of four agents with preferences that contain the following prefixes, respectively:
\begin{itemize}
    \item $v_1:(B,4) \pref_1 (B,3) \pref_1 (A,4) \pref_1 (B,2) \pref_1 (A,3) \pref_1 \cdots$
    \item $v_2: (B,4) \pref_2 (B,3) \pref_2 (B,2) \pref_2 (A,4) \pref_2 (B,1) \pref_2 \cdots$
    \item $v_3: (A,4) \pref_3 (A,3) \pref_3 (A,2) \pref_3 (B,4) \pref_3 (A,1) \pref_3 \cdots$
    \item $v_4: (A,4) \pref_4 (A,3) \pref_4 (B,4) \pref_4 (A,2) \pref_4 (B,3) \pref_4 \cdots$
\end{itemize}

Regardless of how the remainder of the preference profile is filled, as long as monotonicity is maintained, there are at least three stable assignments: (1) all agents at $A$; (2) all agents at $B$; or (3) $v_1$ and $v_2$ at $B$ and $v_3$ and $v_4$ at $A$.
\end{example} 

We would like to identify conditions under which a profile admits a unique stable assignment.
One extreme case is when preferences are non-interleaving.

\begin{definition}[Non-interleaving]
A preference order is \emph{non-interleaving} if it is monotonic and either $(A,1) \pref (B,n)$ or $(B,1) \pref (A,n)$.
A profile is \emph{non-interleaving} if it contains only non-interleaving preference orders.
\end{definition}

The profile in Example \ref{example:stable} is an instance of a non-interleaving profile. 
If an agent's preference is non-interleaving, then their choice of community is independent of the other agents: they would rather be alone at their preferred alternative than with everyone else at the other alternative. Thus, their preference is described by a single bit of information: it suffices
to know whether they are in $V_A^*$ or in $V_B^*$.
Non-interleaving preferences can be viewed as a degenerate case of non-critically-interleaving preferences when $j = 1$.

\begin{observation}
When preferences are non-interleaving, there is a unique stable assignment.
\end{observation}

\begin{proof}
The only stable assignment assigns to all agents in $V_A^*$ to $A$ and all agents in $V_B^*$ to $B$.
Otherwise, an agent assigned to the opposite community will wish to move, even if on their own.
\qed\end{proof}

Non-interleaving preferences can be generalized to domains of preferences that guarantee unique stable assignments. Informally, we say that an agent is $k$-loyal to an alternative $S$ if 
they prefer to be at $S$ with $k$ other agents to being at 
the other alternative in a coalition of size $n$.

\begin{definition}[$k$-Loyalty]
  An agent $v_i \in V$ is {\em $k$-loyal} to alternative $S$, $k\in [n]$, if $v_i\in V_S^*$ and $(S,k)  \pref_i (S',n)$  for $S' \neq S$.
\end{definition}

%
When all agents are sufficiently loyal to their preferred alternatives, there is a unique stable assignment.

\begin{proposition} \label{theorem:allforced}
 Suppose there exist some $k_1$, $k_2$ such that $k_1\leq |V_A^*|$, $k_2\leq |V_B^*|$,
 every agent in $V_A^*$ is $k_1$-loyal, and every agent in $V_B^*$ is $k_2$-loyal.
 Then there is a unique stable assignment.
\end{proposition}

\begin{proof}
By construction, any stable assignment must have all agents in $V_A^*$ assigned to $A$, because otherwise those assigned to $B$ would prefer to move together to $A$, forming a coalition of size $|V_A^*|\ge k_1$ at $A$.
Symmetrically, any stable assignment must have all agents in $V_B^*$ assigned to $B$, as otherwise those assigned to $A$ would prefer to move together to $B$, forming a coalition of size at least $|V_B^*|\ge k_2$ at $B$.
\qed\end{proof}

Proposition~\ref{theorem:allforced} holds because all agents must necessarily be assigned to their preferred alternative.
We now examine a sub-domain of non-critically-interleaving preferences in which there is always a unique stable assignment, but not all agents are necessarily assigned to their preferred alternative.

\begin{proposition} \label{theorem:tworounds}
Suppose agents' preferences are non-critically-interleaving. Let 
\begin{align*}
V_A' &= \argmax\limits_{U \subseteq V} |\{v_i \in U : (A, |U|) \pref (B,n)\}|\ , \\
V_B' &= \argmax\limits_{U \subseteq V} |\{v_i \in U : (B, |U|) \pref (A,n)\}|\ .
\end{align*}
If none of the agents in $V_A^*\setminus V_A'$ are $(n - |V_B'|)$-loyal and none of the agents in $V_B^*\setminus V_B'$ are $(n - |V_A'|)$-loyal, then there is a unique stable assignment.
\end{proposition}

\begin{proof}
Note first that, by monotonicity, the set $\argmax$ in the definition of $V_A'$ and $V_B'$ is a singleton, so $V_A'$ and $V_B'$ are well-defined.
As with Proposition~\ref{theorem:allforced}, for any assignment to be stable it must assign all agents in $V_A'$ to $A$ and those in $V_B'$ to $B$.
For the remaining agents, they must necessarily be assigned to the opposite alternative, because there cannot be enough agents at their most preferred alternative for them to stay there.
\qed\end{proof}

The maximal class of profiles for which there is a unique stable solution is still more general than those we describe above.
We can use Algorithm~\ref{algorithm:general} to characterize the set of profiles that admit a unique stable assignment.
Let $R_A$ be the assignment rule given by Algorithm~\ref{algorithm:general}, and let $R_B$ be the complementary assignment rule that starts with all agents at $B$ and iteratively moves them to $A$ in the same manner.

\begin{theorem}
Algorithm~\ref{algorithm:general} $(R_A)$ and the reverse assignment rule $(R_B)$ return the same assignment if and only if the profile admits a unique stable assignment.
\end{theorem}

\begin{proof}
If the stable assignment is unique, then both $R_A$ and $R_B$ must return this assignment. 
We now show that if $R_A$ and $R_B$ return the same stable assignment, then it must be the unique stable assignment.
Let $V_A^1$ and $V_B^1$ be the communities in the stable assignment $f_1 = R_A(V)$.
Let $V_A^2$ and $V_B^2$ be the communities according to a different stable assignment $f_2$. 
By monotonicity and the properties of Algorithm~1
we have $V_A^1\subsetneq V_A^2$.
Consider the set of agents $V_A^2 \setminus V_A^1$, and in particular, the agent(s) in this set that were the first to be moved to~$B$ by $R_A$. At the beginning of the iteration in which they were moved, the number of agents at $A$ had to be at least $|V_A^2|$ (before moving). This contradicts the claim that $f_2$ is stable, as there are agents in $V_A^2$ preferring to move together to $B$.
\qed\end{proof}

\subsection{Cohesiveness}

Not all stable assignments may be equally attractive. In real life, forking comes at a cost, such as the need to replicate infrastructure and to carve out or abandon intellectual property or goodwill, as well as the social and emotional cost of separation.
The cost of forking within our framework is implicit in the preferences of the agents. In line with the monotonicity of preferences, it is natural that the community may want to avoid forks when possible.
When it is desirable to avoid forking, we prefer stable assignments that place all agents at the same alternative over those that fork. We call these non-forking assignments. 
A profile is said to be \emph{cohesive} if it admits at least one non-forking stable assignment; 
otherwise, we say that a profile is {\em forking}. The profile in Example~\ref{example:manystable} is cohesive, although it also permits a forked stable assignment.

%

The assignment in Example \ref{example:stable} is stable, but the input profile is forking, because no stable assignment exists with all agents in one community.
For a profile to be cohesive there must be at least one alternative (w.l.o.g, $A$) such that for all $j \leq n$, there are fewer than $j$ agents who prefer $(B,j)$ over $(A,n)$.
The following example shows a cohesive profile with no stable forked assignments.

\begin{example}\label{example:nonforkingstable}
Let $V = \{v_1, v_2\}$, where $(A,2) \pref_1 (B,2) \pref_1 (A,1) \pref_1 (B,1)$ and $(B,2) \pref_2 (A,2) \pref_2 (B,1) \pref_2 (A,1)$. Each agent would prefer to be together with the other agent at their less preferred alternative rather than alone at their preferred alternative. This profile is cohesive, and only admits stable assignments that are non-forking.
\end{example}


\section{Strategyproofness}\label{section:strategicbehavior}
So far we have considered the existence and the possibility of efficiently computing stable assignments when agents report their preferences truthfully. Another important question is whether there exist strategyproof stable assignment rules, i.e., rules that output stable assignments and do not incentivize the agents to misreport their true preferences.

\begin{definition}[Strategyproofness]
A rule $R$ is {\em strategyproof over domain $D \subseteq \T^n$} if for all profiles $V \in D$ and assignments $f = R(V)$, there is no agent $v_i \in V$ that can unilaterally change her preference order to $v_i'$, creating a new profile $V'$ such that she prefers $f'(v_i)$ over $f(v_i)$, 
where $f' = R(V')$.
\end{definition}

Similarly, a rule is $k$-strategyproof in our setting if no subset of agents of size $k$ can simultaneously report false preferences to yield an assignment they all prefer. Naturally, $k$-strategyproofness implies $(k-1)$-strategyproofness.

\begin{definition}[$k$-Strategyproofness]
A rule $R$ is {\em $k$-strategyproof over domain $D \subseteq \T^n$} if for all profiles $V \in D$ and assignments $f = R(V)$, there is no subset of agents $U \subseteq V$ of size $|U| \leq k$ that can simultaneously change their preferences, creating a new profile $V'$ such that each agent $v_i \in U$ prefers $f'(v_i)$ over $f(v_i)$, where $f' = R(V')$.
\end{definition}

For the domain of all monotonic profiles, no strategyproof stable rules exist.
This can be seen from Example \ref{example:nonforkingstable}.
In this example there are two stable assignments, one creating $(A,2)$ and the other creating $(B,2)$. Suppose the agents are both assigned to $B$. If $v_1$ were to change their reported preferences to $(A,2) \pref_1 (A,1) \pref_1 (B,2) \pref_1 (B,1)$, then $(A,2)$ would become the only stable assignment for the new profile, which $v_1$ clearly prefers over $(B,2)$. This profile is symmetric, so if the agents were to be assigned to $(A,2)$ (by some tie-breaking mechanism) then $v_2$ has the opportunity to be strategic.
In fact, no strategyproof stable assignment rule can exist for any domain containing a profile that admits two or more stable assignments.

\begin{theorem} \label{theorem:notstrategyproof}
No assignment rule can be strategyproof over a domain that includes a profile that admits more than one stable assignment.
\end{theorem}

\begin{proof}
Suppose profile $V$ permits at least two stable assignments, and our assignment rule~$R$ picks one of them, $f_1 = R(V)$.
%
%
%
Let $f_2$ be the closest stable assignment to $f_1$, in the sense that there is no other assignment $f_3$ such that $|f_3^{-1}(A)|$ is between $|f_1^{-1}(A)|$ and $|f_2^{-1}(A)|$, and consequently no $|f_3^{-1}(B)|$ between $|f_1^{-1}(B)|$ and $|f_2^{-1}(B)|$ (indeed, $f^{-1}(A) + f^{-1}(B) = n$ for any $f$).
For brevity, let $V_A^1 = f_1^{-1}(A)$, $V_A^2 = f_2^{-1}(A)$,
$V_B^1 = f_1^{-1}(B)$, $V_B^2 = f_2^{-1}(B)$.
Assume that $|V_A^1| > |V_A^2|$ and $|V_B^2| > |V_B^1|$; later we will see that this assumption is without loss of generality.

From monotonicity, we know that $V_A^2 \subset V_A^1$ and $V_B^1 \subset V_B^2$.
Consider an agent $v_i \in V_A^1$ with $(B, |V_B^2|) \pref_i (A, |V_A^1|) \pref_i (B, |V_B^1|+1)$. At least one such agent must exist because otherwise $f_2$ could not be stable, as all the agents in $V_A^1 \cap V_B^2$ would prefer to move together to $A$ rather than stay in $(B, |V_B^2|)$.
If $v_i$ commits to $B$ by falsely reporting that they prefer $(B, 1)$ to  $(A,n)$, then they must be assigned to~$B$ (as otherwise the assignment would not be stable). By construction, however, no intermediate stable state can exist between $f_1$ and $f_2$, so agents will prefer to move from $V_A^1$ to $B$ until it is of size $|V_B^2|$, which is what $v_i$ preferred.
By symmetry, if our rule $R$ had picked assignment $f_2$ instead of $f_1$ then we would have the same result; thus, our assumption that $|V_A^1| > |V_A^2|$ and $|V_B^2| > |V_B^1|$ is indeed
without loss of generality.
\qed\end{proof}

The above theorem means that the domain consisting of those profiles that admit a unique stable assignment is the maximal domain for which a stable assignment rule can be strategyproof.
Next, we show that, whenever a profile admits a unique stable assignment, Algorithm~\ref{algorithm:general} is strategyproof.

\begin{lemma} \label{lemma:strategyproof}
Algorithm~\ref{algorithm:general} is strategyproof over the domain of all profiles that admit a unique stable assignment.
\end{lemma}

\begin{proof}
Consider a run of Algorithm~1 in which it assigns some agent $v_i$ to~$B$ and outputs the assignment $f$.
First, observe that $v_i$ cannot misreport their preferences so that they would end up in a larger community at $B$ that they prefer. If the agents assigned to $A$ do not move at any iteration, then $v_i$ moving at an earlier to later iteration has no effect on them. And, due to monotonicity, $v_i$ staying at $A$ would not further entice anyone to move to $B$.

Second, we want to show that agent $v_i$ cannot manipulate the outcome so that it will be assigned to a larger community at $A$ that it prefers.
Suppose that at the beginning of the iteration when $v_i$ is moved from $A$ to $B$, the size of the community at $A$ is $a$.
The agents who were moved from $A$ to $B$ at an iteration before the iteration at which $v_i$ is moved will be assigned to $B$ regardless of what $v_i$ reports.
In general, agents moved to $B$ together at one iteration must end up at $B$ regardless of the preferences of those moved to $B$ at later iterations and those who stay at $A$.
As a consequence, $v_i$ can never induce an assignment with a community at $A$ of size greater than~$a$.
Since $v_i$ was moved at the iteration when the size of $A$ was $a$, it must be to a community $B$ that they prefer over $(A,a)$.
Therefore no agent $v_i \in B$ can deviate profitably from their true preferences.

It remains to show that no agent assigned to $A$ can benefit from strategic behavior. By symmetry, if there is a unique stable assignment, then if Algorithm~\ref{algorithm:general} starts with all agents at $A$ and moves them in batches to $B$, or starts at $B$ and moves them in batches to $A$, then it must return the same assignment.
We can therefore use the same argument as above for agents assigned to $A$ according to the algorithm that initializes $A$ and $B$ in the opposite way.
\qed\end{proof}

The result extends to $n$-strategyproofness, or group-strategyproofness, since, if we consider any coalition of agents assigned to $B$ by Algorithm~\ref{algorithm:general}, and look at only those who were moved first (in the same iteration as one another, but before everyone else in the coalition who was moved), then they have no incentive to misreport their preferences for the same reason as the agent in the proof of Lemma~\ref{lemma:strategyproof}.
By combining this with Theorem~\ref{theorem:notstrategyproof}, we arrive at our main result.

\begin{theorem}
Algorithm~\ref{algorithm:general} is group-strategyproof over the domain of all profiles that admit a unique stable assignment.
\end{theorem}

Just how common are profiles that admit strategyproof stable assignment rules?
One specific domain restriction that implies group-strategyproofness of Algorithm~\ref{algorithm:general} is the domain of non-interleaving profiles (recall that, for this domain restriction, placing all agents at their preferred alternative is stable).
Non-interleaving preferences are indeed very extreme, in that agents ignore each other completely.
However, if we relax this extreme constraint on preferences even the slightest bit, we can lose strategyproofness.

\begin{definition}[Minimally-interleaving]
A preference order is \emph{minimally-interleaving} if it is monotonic and $(A,2) \pref (B,n) \pref (A,1) \pref (B, n-1)$ or $(B,2) \pref (A,n) \pref (B,1) \pref (A, n-1)$.
A profile is a \emph{minimally-interleaving} if it contains only non-interleaving and minimally-interleaving preferences.
\end{definition}

Minimally-interleaving preferences can be interpreted as just barely extending non-interleaving preferences to allow that agents may be willing to go with their less preferred alternative if they would otherwise be alone with their more preferred alternative. Note that the minimally-interleaving domain is still a severely restricted domain. In particular, it allows each agent to specify only one of four possible orders.
%
However, it turns out that if we allow just minimal interleaving, then there is no assignment rule that is both stable and strategyproof.

\begin{observation}\label{theorem:1Inter_strategy}
There is no assignment rule that is both stable and strategyproof for all minimally-interleaving preference profiles. 
\end{observation}

\begin{proof}
Let $R$ be a stable assignment rule and consider the profile from Example~\ref{example:nonforkingstable}:
%
$v_1: (A,2) \pref_1 (B,2) \pref_1 (A,1) \pref_1 (B,1)$;
$v_2: (B,2) \pref_2 (A,2) \pref_2 (B,1) \pref_2 (A,1)$.
%
Note that, indeed, this profile is minimally-interleaving.
Observe that the only stable assignments are the two that place both agents in the same community. Since there are two stable assignments, Theorem~\ref{theorem:notstrategyproof} implies that $R$ cannot be strategyproof for this profile.
For illustrative purposes, consider the case in which $v_1$ votes strategically by reporting $v_1': (A,2) \pref_1 (A,1) \pref_1 (B,2) \pref_1 (B,1)$. 
The only stable assignment places both agents at $A$, creating $(A,2)$, which $v_1$ prefers to $(B,2)$.
So if $R$ placed both agents at $(B,2)$, it cannot be strategyproof. 
\qed\end{proof}

We say that a profile is {\em $k$-interleaving} if it may contain preference orders in which $(S',n) \pref \dots \pref (S,n) \pref (S',k)$, but not in which $(S',n) \pref \dots \pref (S,n) \pref (S',k+1)$, where $S' \neq S$.
Hence, non-interleaving preferences are equivalent to $0$-interleaving; minimally-interleaving preferences are the same as $1$-interleaving; and $n$-interleaving is the domain of all monotonic preferences. Naturally, the set of all $k$-interleaving preferences encompasses all $(k-1)$-interleaving preferences, so no strategyproof stable assignment rule can exist for $k \geq 1$.
While non-interleaving is a sufficient condition for strategyproofness, the next example demonstrates that it is not a necessary condition.

 \begin{example}\label{example:mixed}
 Consider two agents, $V = \{v_1, v_2\}$, where $v_1: (A,2) \pref_1 (A,1) \pref_1 (B,2) \pref_1 (B,1)$ and $v_2: (B,2) \pref_2 (A,2) \pref_2 (B,1) \pref_2 (A,1)$.
 Any stable assignment must have $v_1$ at $A$, independent of the preferences of $v_2$. Agent $v_2$ would prefer to be at $A$ with $v_1$ to being alone at $B$, so the only stable assignment has both agents at~$A$, and neither agent has an incentive to be strategic.
 \end{example}

 Notice that our interleaving conditions apply to the preference order of each agent individually. Example \ref{example:mixed} suggests that we should instead consider restrictions on the profile as a whole.
 While we know that the necessary and sufficient conditions for stable strategyproof assignment rules to exist is that there be a unique stable assignment, characterizing the profiles for which this occurs is an interesting challenge.


\section{Forking with More than Two Alternatives} \label{sec:three}

So far, we focused on the case of two alternatives. We conclude the paper with two observations about the general case:
  (1) stable assignments are no longer guaranteed to exist;
  (2) deciding whether an assignment is stable is NP-complete.

\begin{proposition}
There exist monotonic profiles with no stable assignment.
\end{proposition}

\begin{proof}
Consider the problem with three agents $V = \{v_1, v_2, v_3\}$, three alternatives $\{A,B,C\}$, and a following profile:
\begin{itemize}[topsep=0pt]

\item
$v_1: \cdots \pref_1 (B,2) \pref_1 (A,2) \pref_1 (A,1) \pref_1 (B,1) \pref_1 (C,3) \pref_1 \cdots$

\item
$v_2: \cdots \pref_2 (C,2) \pref_2 (B,2) \pref_2 (B,1) \pref_2 (C,1) \pref_2 (A,3) \pref_2 \cdots$

\item
$v_3: \cdots \pref_3 (A,2) \pref_3 (C,2) \pref_3 (C,1) \pref_3 (A,1) \pref_3 (B,3) \pref_3 \cdots$

\end{itemize}
%

Assume for contradiction that this profile admits a stable assignment $f$.
As $v_2$ prefers $(B,1)$ to  $(A,3)$, the assignment $f$ cannot assign $v_2$ to $A$.
By considering $v_1$ and $v_3$, we conclude that $|f^{-1}(s)|<3$ 
for every $S\in\{A, B, C\}$.
%
%
%
Suppose $f(v_1)=A$, $f(v_3)=A$. Then we have $f(v_2)=B$ 
because $(B,1) \pref_2 (C,1)$. But in this case, $v_1$ would prefer to move to $B$ since $(B,2) \pref_1 (A,2)$.
By the same reasoning $|f^{-1}(S)|\neq 2$ for each $S\in\{A, B, C\}$.
The only remaining option is to have one voter at each alternative. Let $v$
be the voter at $A$. If $v=v_1$, then $v_1$ prefers to move to $B$
and if $v=v_3$ then $v_3$ would prefer to move to $C$. Finally, if $v=v_2$
then $v_1$ prefers to join $v_2$ at $A$.
\qed\end{proof}

From a complexity-theoretic perspective, it is then natural to ask if there are efficient algorithms 
for (a) checking whether a given assignment is stable, and (b) deciding if a given profile
admits a stable assignment. It turns out that, while the answer to the first question is `yes', 
the answer to the second question is likely to be `no'.

\begin{proposition}\label{prop:in-np}
We can decide in polynomial time whether a given assignment for a forking problem is stable.
\end{proposition}
\begin{proof}
Note first that if an assignment $f$ is not stable, then this can be witnessed
by a deviation in which all deviating agents move to the same alternative (say, $A$).
Indeed, the agents who deviate from $f$ by moving to $A$ would find this move beneficial 
even if other agents did not move (in particular, due to monotonicity, they benefit 
from other agents not moving away from $A$). Thus, to decide if a given assignment $f$ is stable, 
it suffices to consider deviations that can be described by a pair $(S, n_S)$,
where $S$ is an alternative and $n_S>f^{-1}(S)$. For each such pair, we need
to check if there are $n_S-f^{-1}(S)$ agents who are currently not assigned to $S$,
but prefer $(S, n_S)$ to their current circumstances.
\qed\end{proof}

\begin{proposition}\label{prop:nphard}
Deciding whether a forking problem admits a stable assignment is NP-complete.
\end{proposition}
\begin{proof}[Sketch]
By Proposition~\ref{prop:in-np}, our problem is in NP. 
For hardness we adapt the reduction argument of Darmann~\cite[Th. 3]{darmann2015group}, establishing NP-hardness for the core stability problem in {\sf o-GASP} with increasing preferences.
That construction makes use of so-called void activities, which are available in {\sf o-GASP}, but not in forking problems. In the profile constructed for~\cite[Th. 3]{darmann2015group}, the occurrence of void activities in each agent's preference needs to be replaced by $(S^*, 1)$, where $S^*$ denotes the top alternative in the agent's preference.
\qed\end{proof}
Our hardness reduction produces an instance where the number of alternatives
is linear in the number of voters. The complexity of finding a stable assignment for a fixed number of
alternatives (e.g., $m=3$) remains open.

\section{Conclusions and Future Work} \label{sec:conclusions}

In the real world, communities sometimes fracture, or fork. This can generally be seen as a consequence of the decisions the community has made. 
If agents associate freely, with the ever-present option of leaving, then we can account for this possibility within collective decision making procedures. This enables minorities to threaten a fork in protest against the tyranny of the majority while giving the majority an opportunity to concede to prevent a fork.
Such a forking process also facilitates the emergence of new communities, as it may be easier to sprout a community from an existing one rather than to build one from scratch.

We have shown that, while it may not be difficult to find stable partitions of a set of agents, constructing strategyproof rules is only possible in restricted domains. While the necessary and sufficient conditions for strategyproofness remain an interesting open question, we have identified a range of circumstances that are sufficient for strategyproofness.
Lastly, we have shown that efficient preference elicitation is possible and desirable.

The social choice setting we considered is, to the best of our knowledge, novel and our work has only made the first steps towards its analysis. Several directions for future research present themselves:
(1) first, as mentioned above, settling the question about the domain restrictions that are necessary and sufficient for the existence of stable and strategy-proof assignment rules is a priority;
(2) second, natural generalizations of the setting we propose will be worth investigating---e.g., settings with several alternatives (similarly to how, e.g., large miners can be present in several forks), or settings in which the identities of the agents matter (as agents may wish to fork with other specific agents); 
(3) third, studying mechanisms for the converse problem, in which several communities could merge into a new one;
and
(4) fourth, enabling a majority to remove troublesome or faulty agents (e.g. Sybils) by forcing a fork.


\section*{Acknowledgements}
Ehud Shapiro is the Incumbent of The Harry Weinrebe Professorial Chair of Computer
Science and Biology. We thank the generous support of the Braginsky Center for the
Interface between Science and the Humanities. Nimrod Talmon was supported by the
Israel Science Foundation (ISF; Grant No. 630/19).
Ben Abramowitz was supported in part by NSF award CCF-1527497.


\bibliographystyle{plain}
\bibliography{bib}

\begin{thebibliography}{10}

\bibitem{tezos}
V.~Allombert, M.~Bourgoin, and J.~Tesson.
\newblock Introduction to the {T}ezos blockchain.
\newblock In {\em Proceedings of HPCS '19}, pages 1--10, 2019.

\bibitem{beck2018governance}
R.~Beck, C.~M{\"u}ller-Bloch, and J.~L. King.
\newblock Governance in the blockchain economy: A framework and research
  agenda.
\newblock {\em Journal of the Association for Information Systems}, 19(10):1,
  2018.

\bibitem{brandt2016handbook}
F.~Brandt, V.~Conitzer, U.~Endriss, J.~Lang, and A.~D. Procaccia.
\newblock {\em Handbook of Computational Social Choice}.
\newblock Cambridge University Press, 2016.

\bibitem{bulteaupartition}
L.~Bulteau, P.~Jain, and N.~Talmon.
\newblock Partition aggregation for budgeting.
\newblock In {\em Proceedings of M-PREF '20 (at ECAI '20)}, 2020.

\bibitem{cgt-book}
G.~Chalkiadakis, E.~Elkind, and M.~Wooldridge.
\newblock {\em Computational aspects of cooperative game theory}.
\newblock Morgan {\&} Claypool Publishers, 2011.

\bibitem{darmann2015group}
A.~Darmann.
\newblock Group activity selection from ordinal preferences.
\newblock In {\em Proceedings of ADT '15}, pages 35--51. Springer, 2015.

\bibitem{darmann2018social}
A.~Darmann.
\newblock A social choice approach to ordinal group activity selection.
\newblock {\em Mathematical Social Sciences}, 93:57--66, 2018.

\bibitem{darmann2018group}
A.~Darmann, E.~Elkind, S.~Kurz, J.~Lang, J.~Schauer, and G.~Woeginger.
\newblock Group activity selection problem with approval preferences.
\newblock {\em International Journal of Game Theory}, 47(3):767--796, 2018.

\bibitem{deliberativemajorities}
E.~Elkind, D.~Grossi, E.~Shapiro, and N.~Talmon.
\newblock United for change: Deliberative coalition formation to change the
  status quo.
\newblock In {\em Proceedings of AAAI '21}, 2021.

\bibitem{Grossi_2014}
D.~Grossi and G.~Pigozzi.
\newblock {\em {Judgment Aggregation: A Primer}}.
\newblock Morgan {\&} Claypool Publishers, 2014.

\bibitem{bitcoin}
S.~Nakamoto.
\newblock Bitcoin: A peer-to-peer electronic cash system.
\newblock Technical report, Manubot, 2019.

\bibitem{narayanan16handbook}
A.~Narayanan, J.~Bonneau, E.~Felten, A.~Miller, and S.~Goldfeder.
\newblock {\em Bitcoin and Cryptocurrency Technologies}.
\newblock Princeton University Press, 2016.

\bibitem{cryptoimpact}
D.~Ng and P.~Griffin.
\newblock The wider impact of a national cryptocurrency.
\newblock {\em Global Policy}, page~1, 2018.

\bibitem{cryptocurrency}
A.~Phillip, JSK Chan, and S.~Peiris.
\newblock A new look at cryptocurrencies.
\newblock {\em Economics Letters}, 163:6--9, 2018.

\bibitem{reijers2018now}
W.~Reijers, I.~Wuisman, M.~Mannan, P.~De~Filippi, C.~Wray, V.~Rae-Looi, A.~C.
  V{\'e}lez, and L.~Orgad.
\newblock Now the code runs itself: On-chain and off-chain governance of
  blockchain technologies.
\newblock {\em Topoi}, pages 1--11, 2018.

\bibitem{liquidgit}
B.~Swierczek.
\newblock Democratic file revision control with liquidfeedback.
\newblock {\em The Liquid Democracy Journal}, 2021.

\bibitem{wattenhofer17distributed}
R.~Wattenhofer.
\newblock {\em Distributed Ledger Technology: The Science of the Blockchain}.
\newblock Createspace Independent Publishing Platform, 2017.

\bibitem{hardfork}
N.~Webb.
\newblock A fork in the blockchain: Income tax and the bitcoin/bitcoin cash
  hard fork.
\newblock {\em North Carolina Journal of Law \& Technology}, 19(4):283, 2018.

\bibitem{zhou2020has}
S.~Zhou, B.~Vasilescu, and C.~K{\"a}stner.
\newblock How has forking changed in the last 20 years? a study of hard forks
  on github.
\newblock In {\em Proceedings of ICSE '20}, pages 445--456. IEEE, 2020.

\end{thebibliography}












\end{document}